\input harvmac

\def\R{\relax{\rm I\kern-.18em R}}
\font\cmss=cmss10 \font\cmsss=cmss10 at 7pt
\def\Z{\relax\ifmmode\mathchoice
{\hbox{\cmss Z\kern-.4em Z}}{\hbox{\cmss Z\kern-.4em Z}}
{\lower.9pt\hbox{\cmsss Z\kern-.4em Z}}
{\lower1.2pt\hbox{\cmsss Z\kern-.4em Z}}\else{\cmss Z\kern-.4em Z}\fi}\
\def\np{{\it Nucl. Phys.}}
\def\pl{{\it Phys. Lett.} }

\def\prl{{\it Phys. Rev. Lett. }}
\def\p{\partial}

\def\CC{{\cal C}}

\def\l{\lambda}

\def\p{\partial}

\def\tr{{\rm tr}}
 \lref\ksw{V. Kazakov, M. Staudacher and T. Wynter, LPTENS-95/9;  
LPTENS-95/56}  
\lref\dkaz{F. David, \np B 257 (1985) 45;
V. Kazakov, \pl B 150 (1985) 282.}
\lref\bk{
 V.Kazakov, \pl  119 A (1986) 140;
D. Bulatov and  V.Kazakov, \pl 186 B (1987) 379.}
\lref\onon{I. Kostov, {\it Mod. Phys. Lett.} A4 (1989) 217;
 M. Gaudin and I. Kostov, \pl B 220 (1989) 200; I. Kostov and M. 
Staudacher, \np B  384 (1992) 459; B. Eynard and J. Zinn-Justin, \np B 386 
(1992) 55.}  
 \lref\kjm{
 V. Kazakov, \np B (Proc. Supp.) 4 (1988) 93;
J.-M. Daul, Q-states Potts model on a random planar lattice,
 hep-th/9502014.}
  \lref\kmul{
 V. Kazakov, {\it Mod. Phys. Lett.} A4 (1989) 2125.}
\lref\Iade{
  I. Kostov, \np B 326 (1989)583;  \np B 376 (1992) 539.}
 \lref\adem{I. Kostov,  \pl  B 297 (1992) 74.}
 \lref\higkos{
S. Higuchi and I. Kostov,   \pl B 357 (1995) 62.}
 \lref\berchar{
 B. Eynard and C. Kristjansen, preprints SPhT/95-068, 
SPhT/95/133.  }
\lref\phil{ P. Di Francesco and D. Kutasov, \np  B 342  (1990) 589}
 \lref\sixauthors{
M. Douglas and S. Shenker, \np B 335 (1990) 635;
E. Br\'ezin and V. Kazakov, \pl B 236 (1990);
D. Gross and A. Migdal, \prl 64 (1990) 127 ; \np B 340 (1990) 333.}
 \lref\mike{
M. Douglas, \pl  B 238 (1990) 176;
M. Fukuma, H. Kawai and R. Nakayama, { \it Int. J. Mod.
Phys. }  A 6 (1991) 1385;
R. Dijkgraaf, H. Verlinde and E. Verlinde, \np B 348 (1991) 435.}
 \lref\Itrieste{
 I. Kostov,  Solvable Statistical Models on Random Lattices,
Proceedings of the "Conference on recent developments in statistical
mechanics and quantum field theory"  (10 - 12 April 1995), Trieste, Italy.}
 \lref\alfak{
J. Alfaro and I. Kostov, to be published.} 
\lref\KMMM{
S. Kharchev, A. Marshakov, A. Mironov and A. Morozov, 
\np B 397 (1993) 339 and references therein.}
\lref\hirota{
R. Hirota, \prl  27 (1971) 1192.}
 \lref\jimiwa{
 M. Jimbo and T. Miwa, "Solitons and infinite dimensional Lie algebras",
{\it RIMS } Vol. 19 (1983) 943-1001.}
 \lref\mehta{
M. L. Mehta, $Random Matrices$  (second edition), Academic Press, New 
York, 1990.}
 \lref\izii{
C. Itzykson and J.-B. Zuber, J. Math. Phys. 21 (1980) 411.}
 \lref\zamsal{ 
P. Fendley and H. Saleur, \np B 388 (1992) 609,
 Al. B. Zamolodchikov, \np B 432 [FS] (1994) 427.}
\lref\trw{ 
D. Bernard and A. LeClair, \np B426 (1994) 534;
 C. Tracy and H. Widom, Fredholm determinants and
the mKdV/sinh-Gordon hierarchies, solv-int/9506006, S. Kakei, Toda Lattice 
Hierarchy and Zamolodchikov conjecture, solv-int/9510006.}
\lref\DS{Drinfeld and V. Sokolov, {\it. Itogi Nauki i Techniki} 24 (1984) 
81,
{\it Docl. Akad. Nauk. SSSR} 258 (1981) 1.}
\lref\DFK{P. Di Francesco and D. Kutasov, \np B 342 (1990) 589.}
 \lref\KW{V. Kac and M. Wakimoto, Exceptional hierarchies of soliton 
equations, {\it
Proceedings of Symposia in Pure Mathematics}, 49 (1989) 191}
  \Title{}
{\vbox{\centerline
{Generalized Hirota Equations  }
\vskip2pt
\centerline{ in  Models of 2D Quantum Gravity
    }}}

\vskip6pt
 \centerline{Jorge Alfaro }

\centerline{{\it    Universidad Cat\'olica de Chile }}
\centerline{{\it Facultad de Fisica, Av. Vicu\~a Mackenna 4860, Santiago, Chile}}
\bigskip

\centerline{Ivan K. Kostov \footnote{$ ^\ast $}{on leave of absence
from the Institute for Nuclear Research and Nuclear Energy,
Boulevard Tsarigradsko shosse 72, BG-1784 Sofia,
 Bulgaria}  }

\centerline{{\it C.E.A. - Saclay, Service de Physique Th\'eorique}}
 \centerline{{\it 
  F-91191 Gif-Sur-Yvette, France}}

\vskip .3in
\baselineskip10pt{ 
We derive a set of bilinear functional equations  of Hirota type
 for the partition functions of the $sl(2)$ related integrable statistical models defined on a random lattice.
These equations are obtained as deformations of the Hirota equations 
for the KP integrable hierarchy, which are satisfied by the partition function of the ensemble of planar graphs.
    }

\bigskip
\rightline{SPhT-96/029}
\Date{April 1996 }

\baselineskip=16pt plus 2pt minus 2pt

\newsec{Introduction}

The formulation of 2D quantum gravity in terms of  random matrix variables 
opened the possibility to apply  powerful nonperturbative techniques as 
the method of orthogonal polynomials \izii\ and led to the discovery of 
unexpected integrable structures 
associated with its different scaling regimes. Originally, a structure 
related 
to the KdV hierarchy of soliton equations  was found  in the continuum 
limit 
of pure gravity \sixauthors. Consequently, the partition functions of the
  $A$ series of models of matter  coupled to 2D gravity  were identified 
as $\tau$-functions of higher reductions of the KP integrable hierarchy 
\mike.  Similar statement concerning the  $D$ series was made in \phil.
    The reductions of the KP hierarchy reflect the flow structure of the 
theory of interacting gravity and matter at distances much smaller than 
the correlation length of the matter fields. On the other hand, there are 
interesting phenomena as the topology changing interactions and, in 
general, all processes involving world manifolds with negative curvature, 
which become important at distances much larger than the correlation 
length of the matter fields.  At extra large distances the fluctuations of 
the matter fields are not important and the flow structure of the theory 
becomes the one of pure gravity.  Therefore, to study the infrared 
phenomena, we need a complementary description in which the matter is 
considered as a perturbation 
 of pure gravity.  The simplest realization of such a description is given 
by the Kazakov's multicritical points of the one-matrix model \kmul,  
where  ``pure" gravity theories defined as specially weighted lattices  were 
shown to be equivalent to theories of gravity coupled to (nonunitary) 
matter fields.   
     
  In this letter we propose a systematic  construction of the theories with 
matter fields  as deformations of the integrable structure of pure 
gravity.  For this purpose we will exploit the microscopic realization of  
the matter degrees of freedom given by the $sl(2)$-related  statistical 
models: the Ising  \bk \ and the     $O({\rm n})$  \onon\ models, the   
SOS and RSOS models and their $ADE$ and $\hat A \hat D\hat E$ 
generalizations \Iade -\adem. Each of these  models can be reformulated 
as a theory  of one or several random matrices with interaction  of the   
form $\tr\ln (M_a\otimes 1+1\otimes M_{b})$.  Such an  interaction   can 
be introduced by means of  a differential operator of  second order ${\bf 
H}$ acting on the coupling constants;   the partition function of the 
model is obtained by acting with the operator $e^ {\bf H}$ on  the  
partition function of one or several decoupled one-matrix integrals.   
 The Virasoro constraints $L_n^a=0$  for each of the one-matrix integrals 
transform  to linear differential constraints $e^ {\bf H} L_n ^ae^ {-\bf 
H}=0$ for the   interacting theory. These constraints are equivalent to  
the loop equations, which have been already derived by other means \Iade\  
and therefore will not be discussed here. The new point  is that the 
Hirota bilinear equations 
that hold for each of the one-matrix integrals become,  upon replacing the 
vertex operators as  $ {\bf V}_{\pm}(z)^a \to e^ {\bf H}  {\bf 
V}_{\pm}^a(z) e^ {-\bf H}$,  bilinear functional equations for the 
interacting theory.  In this way the integrable structure associated with 
pure gravity is  deformed but not destroyed  by the matter fields.

We start with deriving the Hirota equations for the one-matrix 
integral using the formalism of orthogonal polynomials. Then we show how 
these equations generalize for the matrix integrals describing theories of 
matter fields coupled to gravity.  We restrict ourselves to the 
microscopic  description of the theory, but  our construction survives 
without substantial changes  in the continuum limit where the loop 
equations and the bilinear equations are obtained as deformations of the 
Virasoro constraints and the Hirota equations in the KdV hierarchy. Their  
perturbative solution is given by the loop-space Feynman rules obtained   
in \higkos.

   \newsec{Orthogonal polynomials and  Hirota equations in the one-matrix model}

 The partition function of the ensemble of all two-dimensional 
  random  lattices  is  given  by the hermitian $N\times N$  matrix 
integral
 \eqn\hemm{
Z_N[t]\sim  \int dM \exp \Big( {\rm tr} \sum_{n=0}^{\infty} t_n 
M^n\Big) }
or, in terms of the eigenvalues  $\lambda_i,\ 
i=1,...,N$,
  \eqn\partf{
Z_{N}[t]= \int   \prod _{i=1}^N    d\lambda_i \ \exp 
\Big( \sum_{n=0}^{\infty} t_n \lambda^n_i\Big)  \ \prod_{i< j} (\l_i-\l_j) 
^2 .}
  For the time being we restrict the integration in $\lambda$'s to a  finite 
interval $[\lambda_L, \lambda_R]$  on the real axis, so that the measure
\eqn\meas{d\mu_t(\lambda)=d\lambda \exp 
\Big(\sum_{n=0}^{\infty}t_n\lambda^n\Big)}
is integrable for any choice of the coupling constants $t_n$.

    It is   known \KMMM\   that the partition function  \hemm \  
 is a $\tau$-function of the KP 
hierarchy of soliton equations. 
 The global form of this hierarchy is given by 
the Hirota's   bilinear equations \hirota \ (for a review on the 
theory of the $\tau$-functions see, for example, \jimiwa.)
Below we will  derive the  Hirota equations  using 
 the formalism of    orthogonal 
polynomials.

Introduce for each  $N\ \   (N=0,1,2,...)$ the polynomial
\eqn\polN{\eqalign{
P_{N,t}(\l)&= \langle \det (\l -M)\rangle  _{N, t} \cr
&= {1\over Z_N}  \int \prod _{i=1}^N d\mu_t(\lambda_i) (\l-\lambda_i) \ \prod_{i< j} (\l_i-\l_j) ^2. \cr 
 }}
  It is easy to prove \izii \ that the polynomials \polN\  are orthogonal with 
respect to the measure $d\mu_t(\lambda)$.  
   Indeed,   
   \eqn\otT{\eqalign{
  & Z_N \int d\mu_t(\lambda_{N+1}) P_ {N,t}(\lambda_{N+1})\lambda_{N+1}^k  
\cr
 & \int \prod_{i=1}^{N+1}d\mu(\l_i) \Delta_{N+1}(\l_1,...,\l_{N+1})
  \Delta_N(\l_1,...,\l_N)\l_{N+1}^k\cr  
  &={1\over N+1}\int \prod _{i=1}^{N+1}d\mu_t(\l_i) 
\Delta_{N+1}(\l_1,...,\l_{N+1})\cr
&\sum_{s=1}^{N+1} (-)^{N+1-s}\l^k_s \Delta_{N+1}(\lambda_1,...,\hat 
\l_s,...,\l_{N+1}).\cr}}
 The sum in the integrand is the expansion of the determinant
 $$\left |
 \matrix{
 1 &\l_1& \cdots &\l_1^{N-1}&\l_1^k\cr
 1 &\l_2& \cdots &\l_2^{N-1}&\l_2^k\cr
  \cdots &\cdots     & \cdots  &\cdots &\cdots     \cr
 1 &\l_{N+1}& \cdots &\l_{N+1}^{N-1}& \l_{N+1}^k\cr
 }
 \right | $$
 with respect to its last column. It vanishes for $k=1,..., N-1$, which 
 proves the statement. For $n=N$, one finds
 \eqn\normp{Z_N\int d\mu_t(\l) P_{ N,t}(\l) \l^N= Z_{N+1}/(N+1).}
 Hence
  \eqn\Ort{
\int_{\lambda_L}^{\lambda_R}  
d\mu_t(\lambda)  \   
P_{ N,t}(\lambda)   P_{k,t}(\lambda)=\delta_{N,k} 
\ {Z_{N+1}[t] 
\over (N+1) Z_N[t]} . 
  }

The orthogonality   relations \Ort\ can be  written in  the form of a  
contour integral, 
namely, 
 \eqn\dDrd{
{1\over 2\pi i}\oint _{\cal C} dz\  \Bigg\langle \det 
(z-M)\Bigg\rangle_{ N,t}
\Bigg\langle {1\over \det (z-M)}\Bigg\rangle_{[k+1, t]}=\delta_{N,k}
   }
    where the integration contour $\CC$ encloses the point $z=0$ and the 
interval  
$[\lambda_L,\lambda_R]$.
Indeed,  the residue of  each of
 the $n$ poles is 
equal 
 to the left hand side of  \Ort \  multiplied by 
 $  Z_{k}[t] /  Z_{k+1}[t] $.

A set of more powerful  identities  follow from the fact that   the 
polynomial  $P_{N,t}(z)$
 is orthogonal to $any$ polynomial of degree less than $N$ and in 
particular to the  polynomials $P_{k,t'}[\l], k=1,2,...,N-1$
where $t'= \{ t'_n, n=1,2,...\}$ is another set   of coupling constants. 
Written in the form of contour integrals, these orthogonality relations 
state, for  $N'\le N$,
  \eqn\hirotd{
 \oint _{\cal C} dz\ e^{ \sum_{n=1}^{\infty}
(t_n-t'_n)z^n}
 \Bigg\langle \det (z-M)\Bigg\rangle_{ N,t}
\Bigg\langle {1\over \det (z-M)}\Bigg\rangle_{N', t'}=0 . 
    }
 
The Hirota equations  for the KP hierarchy are obtained from  eq. 
 \hirotd \  after expressing the mean values of $\langle \det^{\pm 
1}\rangle $ in terms of     the vertex 
operators
 \eqn\vop{
{\bf V}_{\pm}(z)= \exp\Big(  \pm \sum_{n=0}^\infty t_n z^n \Big) 
\exp\Big( \mp \ln {1\over z} \ {\partial\over \partial t_0}
\mp 
\sum_{n=1}^\infty 
{ z^{-n}\over n} 
 {\partial\over \partial t_n} \Big).
   }
It follows  from the     definition   \partf \ 
that  
\eqn\vtop{
{\bf V}_\pm(z)\cdot Z_N [t]
 =    e^{\pm \sum_{n=0}^\infty t_n z^n} \Big\langle  \det (z-M)^{\pm 
1}\Big\rangle_{ N,t} Z_N[t]
  }
 and eq.  \hirotd \ is therefore equivalent  to
\eqn\hireqn{
\oint_{\cal C}  dz \   \Big({\bf V}_+(z)\cdot Z_N[t] 
\Big)\ 
\Big({\bf V}_-(z) \cdot Z_{N'}[t']\Big) 
=0  \ \ \ \ \ \ (N'\le N ),
   }
    which is one of the forms of the Hirota equation for KP \jimiwa .
   
    After a change of variables  
\eqn\hireqys{
x_n= {t_n+t'_n\over 2}, \ \ y_n= {t_n-t'_n\over 2},
 } 
the Hirota equations  \hireqn \ take its  canonical form  
\eqn\hirver{
  {\rm Res}_{z=0}  \ z^{N-N'}
 e^{   \sum _{n=1}^{\infty}  2y_n z^n }e^{-\sum 
_{n=1}^{\infty}
  {z^{-n}\over n} {\partial\over \partial y_n} } Z_N[x+y]Z_{N'}[x-y]=0}
  where $ N'\le N$.
The differential equations of the KP hierarchy are obtained by expanding 
 \hirver \ in $y_n$.    
For example, for $N'=N$,  the   
coefficient in front of $y_1^3$  is 
\eqn\olki{
\Big({\partial^4\over\partial y_1^4}+3{\partial^2\over\partial y_2^2}-4
{\partial\over\partial y_1}{\partial\over\partial y_3}\Big)Z_N[t+y]
 Z_{N}[t-y]\Big|_{y=0}=0
  }
and one finds  for the ``free energy'' $u[t]= 2 
{\partial^2\over\partial t_1^2}
\log Z_N$ the KP equation
\eqn\KPe{
3{\partial^2 u\over\partial t^2_2} + {\partial\over\partial t_1}
\Big[ -4 {\partial u\over\partial t_3} +6 u {\partial u\over\partial t_1}
+{\partial^3 u\over\partial t_1^3}\Big]=0.
  }
 The lowest equation in the case   $N'=N-1$ (modified KP) 
 is the so called Miura transformation 
relating the 
functions $u[t]$  and $v[t]= \log (Z_{N+1}[t]/Z_N[t])$: 
  \eqn\Me{u=
 \partial_2 v-\partial_1^2 v -(\partial_1v)^2 .
  }
   
 \newsec{ Bilinear functional equations in the $O({\rm n})$ model}
The partition function of the  $O({ \rm n}) $ matrix model\foot{ We use a  
Roman 
letter for the parameter n$\in[-2,2]$  to avoid confusion with the  
summation index $n$ running the set of natural numbers.}    is defined by 
the  $N\times N$  matrix integral \onon
 \eqn\mort{
 Z_N ^{ O({\rm n})} [t ]\sim 
 \int dM  \exp\Big[ \sum_{n=0}^{\infty}
 t_n \ {\rm tr}M^n 
+ {{\rm n}\over 2}
 \sum_{n+m\ge 1 } {T^{-n-m}\over n+m} \
 {(n+m)!\over n!\ \ m!} {\rm tr}M^n 
 {\rm tr}M^m 
 \Big].
  }
   (the sum over $n$ and $m$ runs the set of nonnegative integers) and  
describes the  ensemble of nonintersecting loops on a 
random graph. A loop of length $n$  is weighted by a factor  ${\rm  {n}} 
T^{-n}$ where 
  $T$ is the   temperature of the loop gas.
 For  n$\in [-2,2]$, the model exhibits    critical behavior  with 
spectrum of the central charge   $C=1-6(g-1)^2/g$,  $g=  {1\over \pi } 
\arccos (-{\rm n}/2)$;   the critical behavior for $ |{\rm n}|>2$ 
is this of a branched polymer \berchar .
   The  matrix integral  \mort \ 
can be again interpreted as the partition function of a Coulomb gas with 
  more complicated but still pairwise  interaction between particles:
\eqn\pfon{
 Z_N ^{ O({\rm n})} [t ]= \ T^{-{1\over 2}{\rm n}N^2} \int   
\prod _{i=1}^N    {d\lambda_i \ e^{\sum_{n} t_n \lambda^n_i} \over (T-2\l_i)^{{\bf\rm n}/2}}
\    \  \prod_{i<j} {(\lambda_i - \lambda_j)^2 \over
 (T-\lambda_i-\lambda_j )^{{\bf\rm n}}}. 
   }
 The   interval of integration
$[\lambda_L, \lambda_R]$   should be such that   $\lambda_R\le T/2 $.  With 
the last restriction the  denominator in the integrand never vanishes.
 The choice of the integration interval does not influence the quasiclassical
 expansion and hence the geometrical interpretation in terms of a  gas 
 of loops on the random planar graph.  
 The saddle point spectral density  is  automatically supported by an
  interval on the half line
  $[-\infty, {T\over 2}]$ \onon.  The most natural choice for the eigenvalue interval 
  is therefore $\l_L\to-\infty, \l_R\to T/2$, with the conditions 
  ${d\mu(\l)\over d\l}|_{-\infty}={d\mu(\l)\over d\l}|_{T\over 2}=0$.

The   $O({\rm n} )$ model reduces to the hermitian matrix model in the 
limit n$\to 0$ and/or $T\to\infty$, and 
can be considered as a deformation of the latter in the 
following sense.
Let us define  the differential operator
\eqn\perth{
{\bf H}=   { {\rm n}\over 2} \Bigg[ -\ln {1\over T}\ {\partial^2\over\partial t_0^2} 
+\sum_{n+m\ge 1 } {T^{-n-m}\over n+m} \
  {(n+m)!\over n!\ \ m!} \ {\partial\over\partial t_n}
{\partial\over\partial t_m} \Bigg] 
}
acting on the coupling constants. It is easy to see that 
  the  partition function \pfon\ is obtained from the 
partition function of the one-matrix model by acting with the operator
 $e^{\bf H}$:
  \eqn\oppq{
   Z_N ^{ O({\rm n})}  [t]=  e^ {\bf H} \cdot Z_N[t] .
}
This simple observation will be of crucial importance for our further  
consideration. It means that  the integrable 
structure  of the 
one-matrix model survives in some form in the 
 $O(n)$ model.   
   
    The   Hirota  equations  
 \hireqn\ provide, due to  the relation \oppq 
,  a set of  bilinear equations   for the 
partition function of the 
$O({ \rm  n})$ model
 \eqn\hireqo{{
 \oint _{\cal C_-} dz   \Big( 
  \tilde {\bf V}_+ (z)\cdot   Z _N^{ O({\rm n})} [t] 
\Big)\ 
\Big( \tilde  {\bf V}_-  (z)\cdot    Z _{N'}^{ O({\rm 
n})}[t']\Big) 
=0  \ \ \  \ \ \  
(N'\le N) 
}}
 where 
the integration contour $C_-$ encloses the interval $[\l_L,\l_R]$ and leaves outside point $z$ and 
the interval $[T-\l_R, T-\l_L]$, and 
$ \tilde {\bf V}_{{\pm} }   (z) = e^{ {\bf H}} {\bf V} _{\pm}(z) e^{-\bf H}$
are the transformed   vertex operators whose explicit form is
 \eqn\tvop{\eqalign{\tilde {\bf V}_{{\pm} }    = 
 &
 e^{ \pm \sum_{n=1}^{\infty} t_n z^n } \exp\Big(\mp\Big[  \ln  {1
\over z}  {\p\over\p t_0} + \sum_{n=1}^{\infty} z^{-n} {\p\over \p t_n} \Big]\Big)\cr
&( T- 2z)^{-{\rm n}/2}\exp \Big( \pm{\rm n} \Big[ \ln {1\over  (T-z)}
   {\p\over\p t_0}   + \sum_{n=1}^{\infty} 
  { (T-z )^{- n} 
\over n} \ {\p\over \p t_n} \Big]  \Big).
  \cr} 
 }
  After being 
expanded in $t_n-t_n' $, eq.  \hireqo \ generates   a hierarchy 
   of differential equations, each of them  involving derivatives with 
respect to an infinite
    number of ``times" $t_n$.

The functional relations \hireqo \
  are equivalent to the   bilinear  relations 
\eqn\hiron{\eqalign{
 & \oint_{\cal C_-} {dz\over (T-2z)^{\rm n}} \exp\Big(\sum 
_{n=1}^{\infty}
  	(t_n-t'_n) z^n \Big) \cr
&  	\Bigg\langle   {\det(z-M) \over
\det(T-z-M)^{\rm n}} \Bigg\rangle_{ N,t}  \Bigg\langle
 {  \det (T-z-M)^{\rm n}\over
\det(z-M)}\Bigg\rangle_{[N',t']}=0 \ \ \   \ \ \ \ \ (N'\le N)\cr}
 }
  where   $\langle \ \ \rangle_{ N,t}$ denotes  the average  
corresponding to the partition function  \pfon . 
 These relations can be also
proved directly by  exchanging the order of the integration in the $\l$'s 
and the  contour integration in $z$.

       \newsec{ Bilinear functional equations in the  $ADE$ and $\hat A\hat D\hat E$ models}

    The   $ADE$ and  $\hat A\hat D\hat E$  matrix
      models   give a nonperturbative microscopic realization of the 
      rational string theories with $C\le 1$.
     Each one of these models is associated with a rank $r$    classical 
simply laced  
     Lie algebra (that is, of type  $A_r, D_r, E_{6,7,8}$) or its afine 
extension, and represents a system of $r$ coupled random matrices.  The 
matrices
      $M_a$ of size   $N_a\times N_a \ (a=1,2,...,r)$ are associated 
with     the nodes of the Dynkin  diagram of the simply laced Lie algebra, 
the    latter being  defined by   its adjacency matrix $G$ with elements 
  \eqn\adJ{
G^{ab}=\cases{1 & if the two nodes are the extremities of a link    $<ab>$ 
\cr
0& otherwise.\cr}
 } 
The partition function $Z^{  G}_{\vec N}    [ \vec t]$ depends on $r$ sets 
of coupling constants $\vec t = \{t_{n}^a\ |\ a=1,...,r;\ 
n=1,2,...\}$. 
The interaction is of
nearest-neighbor type and the   measure is a product of factors associated 
with the nodes $a$ and the    links $<ab> $ of the Dynkin 
diagram
\eqn\DdD{\eqalign{ 
Z^{  G}_{\vec N}    [ \vec t] \sim  & \int \prod_{a=1}^r dM_a \exp 
\Bigg(
 \sum_{a=1}^r \sum_{n=1}^{\infty}  t_{n}^a \ \tr M_a^n
\cr & + {1\over 2} \sum_{a,b} G^{ab}   \sum_{m,n=1}^{\infty} 
{T^{-m-n}\over m+n} 
{(m+n)!\over 
m!\ n!}\ \tr M_a^m \ \tr M_b^n \Bigg). \cr}
 }
  
 The target space of the $A_r$ model is an open chain of $r$ points and 
its critical regimes describe  theories  of  $C=1-6{(g-1)^2\over g}$ matter coupled to gravity,  
$g= 1\pm {1\over r+1}  + 2m, \ m\in \Z_+$.  
 The target space of the  $\hat A_{r-1} $ model is a circle with $r+1$ points and 
 its   continuum limit  describes  a  compactified gaussian  field coupled to gravity. 
 The radius of compactification is  $r$  in a  scale where the  self-dual radius is $r=2$.  
 In this sense the  $O(2)$ model can be referred to as the  $\hat A_0$
 model of  the $\hat A$ series.

 Again, the   only nontrivial integration is with respect to the  
eigenvalues 
$\lambda_{ai}  \ ( i=1,...,N_a)$ of the 
matrices $M_a$:
 \eqn\mtriaa{ 
Z^{G}_{\vec N}[\vec t] = \prod _{ a=1}^r\prod_{i=1}^{N_a} d \lambda_{ai}  
 e^{\sum_n t_{n}^a \lambda_{ai }^n } { { \prod_a \prod_{i< j} 
 (\lambda_{ai }-\lambda_{aj} )^2 \over \prod_{<a b>}
   \prod_{i,j}  (T- \lambda_ {ai} -\lambda_ {bj} )}}.
  }  
  The  domain of  integration  is assumed to be   a  
compact  interval $[\lambda_L, \lambda_R]$ with $\lambda_R\le T/2$.

The partition function 
 \mtriaa \ 
can be obtained by acting on 
 the product of $r$ one-matrix partition functions  $  Z_{N_a}[t^a], 
a=1,...,r,$
with the  exponent of the second-order differential operator  
\eqn\jJj{
 {\bf H} = {1\over 2} \sum_{a,b}G^{ab}
   \Bigg[   \ln T^{-1}  {\partial\over \partial t^a_{0}}
 {\partial\over \partial  t^b_{0}} +
\sum_{n+m\ge 1 } {T^{-n-m}\over n+m} \
 {(n+m)!\over n!\ \ m!}  {\partial\over \partial t_{ n}^a}
 {\partial\over\partial t^{ b}_m}\Bigg] ,  
 } 
 namely,
\eqn\xprd{
  Z^G_{\vec N}    [ \vec t]= e^{{\bf H}} \cdot \prod_{a=1}^r 
Z_{N_a}[t^a].
       }
       
           Each of the one-matrix partition functions on the  right hand side of 
  \xprd \
  satisfies the Hirota equations   \hireqn . As a consequence, the  left 
hand 
  side satisfies  a set  of $r$  
functional equations associated with the nodes $a=1,...,r $ of the Dynkin 
diagram.
Introducing the transformed vertex operators
 \eqn\gmma{
    \tilde {\bf V }_{\pm}^{a}(z) 
 =e^{{\bf H}} {\bf V }_{\pm}^{a}(z)  
  e^{{-\bf H}} 
    }
 we find, for each node  $a=1,...,r,$
 \eqn\hireqoa{
 \oint _{\cal C_-} dz   \Big( 
\tilde {\bf V }_+^{a}(z)\cdot  Z _{\vec N}^{G}[\vec t] \Big)\ 
\Big( \tilde {\bf V }_-^{a}(z)\cdot  Z _{\vec N'}^{G}[\vec t']  \Big) =0 \ 
\ \  \ \ \ 
(N'_a\le N_a)
 }
 where the 
  integration contour
${\cal C_-}$ in 
 \hireqoa \ encloses the interval $[\lambda_L, \lambda_R]$ but leaves 
outside the interval $[T-\lambda_R, T-\lambda_L]$.

 Inserting \jJj\  in the definition  \gmma\ we 
 find the explicit form
  of the  dressed vertex  operators    
\eqn\mvoA{\eqalign{
  & \tilde {\bf V }_{\pm}^a(z)  = \cr
& \prod_{b} (T-2z)^{-{1\over 2}G^{ab}} e^{ \pm \sum_{n=0}^{\infty} 
t^a_n z^n}
 \exp \Bigg(\mp\Bigg[ \ln z^{-1}  {\partial \over 
\partial t^a_0}
+  \sum_{n=1}^{\infty} {z^{-n}\over n}{\partial \over 
\partial t^a_n}\Bigg]\Bigg) \cr 
& \exp \Bigg(\pm \sum _b G^{ab} \Bigg[  \ln (T-z)^{-1} {\partial \over 
\partial t^b_0}
   +  \sum_{n=1}^{\infty} { (T-z )^{- n} \over n} 
{\partial \over \partial t^b_n} \Bigg]
 \Bigg) \cr}
}
  and write   the   generalized Hirota equations \hireqoa \     in terms 
of  correlation functions of 
determinants:  
     \eqn\hironA{\eqalign{ 
 &\oint_{\cal C} dz  
{e^{\sum_{n=1}^{\infty} [t^a_n -t'^a_n ]z^n  }\over
 \prod_{b} (T-2z)^{  G^{ab} } }
  	\Bigg\langle   {\det (z-M_a)\over
\prod_b  \det (T-z-M_b)  ^{G^{ab}}} \Bigg\rangle_{\vec N,\vec t}  \cr
& \Bigg\langle
{   \prod_b   \det(T-z-M_b)^{ G^{ab}}\over\det (
z-M_a)}\Bigg\rangle_{[\vec N',\vec t']}=0\ \ \ \ \ \ \ \ \ \ \  \ \ \ 
(N_a'\le N_a) . \cr}}
  Here $\langle \ \ \rangle_{\vec N,\vec V}$ denotes  the average in the 
ensemble 
  described by the partition function  \mtriaa  .  
  It is possible to prove eq.  \hironA \   directly by 
  performing the contour integration. It is essential for the proof  that 
 the interval $[T-\lambda_R, T-\lambda_L]$ is outside  the contour 
$\cal C$.
  
\newsec{Concluding remarks}

We proposed  an alternative   nonperturbative formulation  of 2D gravity 
suited for investigating the infrared properties of the theory.
We used the fact that in the microscopic realization based on    the $sl(2)$ integrable statistical models, 
the matter  can be introduced as  deformation of pure gravity.
As a consequence, the integrable structure of pure gravity is not destroyed 
and manifests itself in the form of bilinear functional equations of Hirota type.
The main difference between the loop equations and the bilinear equations  
is that the first
relate correlation functions boson-like quantities (traces) while the second relate correlation functions 
of fermion-like quantities (determinants).   
  
The continuum version of these equations will be considered elsewhere.  They have the same  form 
as the equations in the microscopic theory,  the only difference being that  the vertex operators 
in the continuum theory are expanded in the half-integer powers of the shifted and rescaled variable $z$. 
The corresponding  coupling constants  are related to the coupling constants 
associated with scaling operators as the coefficients in the Laurent expansions of the same function at two different points: near the edge of the eigenvalue interval and near infinity.  

      \smallskip
{\bf Acknowledgements}

We thank F. David for critical reading of the manuscript.
Part of this work was completed during the  visit of one of the authors (I.K.) at  PUC, which  
 was  part of the program  for cooperation between CNRS (France) and CONICYT (Chile). The work 
of J.A. has been partially supported by Fondecyt 1950809.
  
  \listrefs
  \bye